Granddaughter and voting for a female candidate

# Eiji Yamamura

Seinan Gakuin University

Abstract

This study examines the influence of grandchildren's gender on grandparents' voting behavior using independently collected individual-level data. The survey was conducted immediately after the House of Councilors election in Japan. I observed that individuals with a granddaughter were more likely to vote for female candidates by around 10 % than those without. However, having a daughter did not affect the parents' voting behavior. Furthermore, having a son or a grandson did not influence grandparents' voting behavior.

This implies that grandparents voted for their granddaughter's future benefit because granddaughters may be too young vote in a male-dominated and aging society.

*JEL classification: D72, J12, and J16.*
Keywords: gender difference, grandchild-grandparents relation, voting behavior, election, female candidate.

1. INTRODUCTION

Many studies have provided evidence about the effect of interactions between family members on preference formation and human behaviors. A working mother leads her sons to maintain a positive view on female labor participation and accepting of working women (Fernandez et al., 2004; Kawaguchi & Miyazaki, 2009). Conversely, the influence of children on their parents was also observed. The presence of daughters leads fathers to hold a view of improving women's social status (Glynn & Sen, 2014; Milyo & Schosberg, 2000; Oswald & Powdthavee, 2010). It is observed that daughters provide parents motivation to address social issues from a long-term viewpoint[1]. Having a daughter increases a congressperson's probability of voting liberally to improve women's rights (Washington, 2008). In rapidly aging societies in developed countries, researchers have increasingly devoted great attention to the relationship between grandparents and their grandchildren. Therefore, this study analyzes whether the granddaughter-to-grandparent relationship reduces the gender gap through voting behavior in elections.

Grandparents are anticipated to be instrumental caregivers for their grandchildren because of the overlap between grandparents' and their grandchildren's lives (Greenfield, 2011; Arpino & Bordone, 2017; Boca et al., 2018; Feng & Zhang 2018)[2]. Parents and grandchildren spend time together to form intimate relationships. The subjective well-being of grandparents increased by having a grandchild (Powdthavee. 2011; Dunifon et al., 2020; Wang et al. 2019)[3].

Parenting daughters stimulated their father's attitudinal shift regarding gender norms (Borrel-Porta et al., 2019), which influences voting behavior (Washington, 2008).

Previous studies have investigated the factors that increase the probability of female candidates winning elections (Milyo & Schosberg, 2000; King & McConnell, 2003; Lublin & Brewer, 2003; Crowley, 2006; Hogan, 2007; Bauer, 2020; Moehling & Thomasson, 2020). However, no study has addressed the granddaughter-to-grandparent relationship to consider the outcome of an election. Therefore, this study extends the parent-child relationship to the grandchild-grandparents relationship to consider how the young generation influences the older generation.

I independently collected individual-level data through an Internet survey directly after the 2016 House of Councilors election in Japan. In the survey questionnaire, respondents were asked whether they voted for female candidates. I further inquired about the family structure, the number of daughters and sons, as well as the number of granddaughters and grandsons. Applying this data, I observed that having granddaughters lead their grandparents to vote for female candidates in the election. The contribution of this study is to bridge family economics and voting behavior to provide new evidence that the gender of grandchildren is a key factor in improving female social and economic status in Japan, where female status is lower compared to other developed countries.

The remainder of this paper is organized as follows. In Section 2, I propose testable hypotheses. In Section 3, I describe the setting of Japan and the related data. Section 4 describes the empirical methodology. Section 5 presents the estimation results and interpretations. The closing section provides some reflections and conclusions.

2. HYPOTHESIS

As political leaders, female candidates are generally anticipated to advance the interests of women and girls and improve female social position and labor participation (Priyanka, 2020). Baskaran and Hessami (2018) noted that female candidates receive more preferential votes if a female mayor has recently been elected to office[4]. Female political leaders are less likely to practice corruption that is rampant in a male-dominated society (Brollo & Troiano, 2016).

Female candidates were nominated by their parties in disadvantageous positions on the ballot (Esteve-Volart & Bagues, 2012). However, changes in social norms reduce the gap between male and female candidates (King & Leigh, 2010). Parties tend to nominate female candidates to poorer positions on the ballot (Esteve-Volart & Bagues, 2012). Systematic bias against female candidates has not yet been observed (Frederick & Streb, 2008). After the introduction of a 50% gender quota in candidate lists, voters became more likely to elect female candidates (Bonomi et al., 2013).

Parents of daughters are considered to be satisfied with the improvement of their social position (Oswald & Powdthavee, 2010). As is observed in the relationship between daughter-parent relations, parents of daughters are more likely to vote for policies that favor women and girls (Washington, 2008). Accordingly, I propose *Hypothesis 1*.

*Hypothesis 1: Having a daughter leads her parents to vote for female candidates in an election.*

Existing studies have analyzed the underlying structural changes in political systems from a historical viewpoint. Acceptance of women's suffrage resulted in the allocation of more resources to education and child health (Carruthers & Wanamaker, 2015; Miller, 2008). For instance, Miller (2008) found that women who garnered votes reduced child mortality. Carruthers and Wanamaker (2015) evidenced that female suffrage increased public school expenditures. In other words, compared to their male counterparts, female candidates place more importance on the interests of children who are not yet eligible to vote. Hence, older individuals who have grandchildren are expected to vote for female candidates to reflect the interest of future generations in policymaking because grandparents have deep affection for their grandchildren.

In the case of respondents with granddaughters, granddaughters are less likely than daughters to be entitled to vote for pursuing their own interests because they are too young. Grandparents of granddaughters have a greater behavior intention to increase women's benefits compared to the parents of daughters. Thus, I propose the following *hypothesis 2:*

*Hypothesis 2: The effect of having a granddaughter on her grandparents' voting behavior is greater than the effect of a daughter.*

3. THE SETTING AND DATA

**3.1. Setting of Japan**

Japan Cabinet Office (2019) reported the problem of aging society in the following manner: In Japan, total fertility rate declined from around 3.50 in 1950 to 1.43 in 2017. Meanwhile, in the United States, the rate declined from 3.50 in 1950 to 1.76 in 2017. Compared with other developed Western countries, Japan experienced a more rapid population aging. The rate of population aged–15-64 is nearly 60%, whereas the rate of population over 65 is around 30% in 2017. Naturally, women are expected to work because to avoid a shortage of labor. However, working women encountered difficulties in childrearing because the supply of nursery schools is not adequate to meet demand; therefore, childcare support measures should be enhanced. In addition, it is important to develop and devise countermeasures to combat declining birthrates. For instance, promotion of work style reform for men and women, work-life balance, and women's advancement in the workplace.

According to the Global Gender Gap Index 2020 rankings, Japan was ranked 121st among 153 countries (World Economic Forum 2020)[5]. This evidently reflects the outdated social and economic structures. To address the problems of an aging society, as political leaders, females are expected to be pivotal in sustaining a society under the unprecedented graying population, thus, an improvement in female social position is required to take countermeasures. Due to globalization, international pressure has led to structural changes in Japanese society. In addition to international criticism of the large gender gap in Japan, Japanese women influenced surrounding men to support female involvement in society[6].

**3.2. Data**

To analyze voting behavior in Japan, individual-level data was collected through a web-based survey in July 2016, conducted directly after the House of Councilors election in Japan. I commissioned the Nikkei Research Company to conduct an internet survey. The survey was openly posted on the Nikkei Research webpage, and was conducted until a sufficient sample had been collected. Since the objective was to collect over 10,000 observations, the survey was active until 10,000 observations were collected. Approximately 12,000 respondents were approached to complete the questionnaire. According to a 2015 survey on information technology, over 90% of the working-age population in Japan are web users. Therefore, selection bias for Internet users does not need to be considered.[7]

In this study, I limited samples according to respondents' ages over 40, 50, and 60 because younger respondents were less likely to have grandchildren. In Japan, there are 47 prefectures, which are considered equivalent to election districts. The participants' residential prefectures were asked to identify the election districts where they voted. Of the 47 prefectures, there were no female candidates in the 15 prefectures in the 2016 election. To appropriately estimate voting behavior, the sample is limited to prefectures where female candidates contested in the 2016 election. Further, respondents who did not vote were eliminated from the sample.

Consequently, observations were reduced to 3,843, 2,359, and 937 if respondents were over 40, 50, and 60 years, respectively. In the alternative sub-sample, it was further limited to those who had children over the age of 30 because adults over 30 years are more likely to have a child. Thereafter, the sub-sample size was further reduced. In the Heckman-type two-step probit model, the sub-sample was limited to respondents having children. In the first step, respondents who had grandchildren were selected from those who had children. The size of the sub-sample was 5,025 in the first step, while the size was reduced to 572 in the second step.

Unfortunately, the age data of the grandchildren are not available. However, I can predict the ages of the respondents' grandchildren. In the sample used in this study, the average age of respondents who have grandchildren was 63 years. In 1980, 36 years ago, they were about 27 years old. The following statistics are based on official data (Ministry of Health, Labour and Welfare, 1980, 2017). In 1980, the average age of women with their first child was approximately 26. Therefore, their first child would be around 36 years old when the surveys were conducted. In 2016, the average age of women with their first child was approximately 30. Overall, the average age of grandchildren was approximately 6. At least, grandchildren are likely to be below 18 years of age[8]. Therefore, most grandchildren are predicted not to be entitled to vote in the election.

The questionnaire included questions about whether they voted for female candidates in the election. Apart from the variables described above, I gathered basic economic and demographic data such as sex, age, educational background, parental educational background, household income, job status, marital status, and number of siblings. Table 1 provides definitions of the key variables and their descriptive statistics, based on the sample of respondents over 40 years used in the baseline estimations in columns (1) and (4) of Table 4. Mean value of VOTE FEMALE is 0.29 which indicates 29% voted for female candidates. Similarly, the rate of respondents having granddaughters was 11%, which is the same as the rate of those having grandsons. This suggests that the genders of grandchildren are less likely to suffer grandparents' preference for grandchild gender. The mean age of the respondents was 53.5 years. The mean age of respondents with children was 54.7 years although it is not reported in Table 1. Hence, ages are almost the same for those who do and do not have children. The mean age of the youngest child was 21.6 years, indicating that older children are older comparatively. Considering it jointly that respondents were, on average, 33 years when they had the youngest child.

Table 2 compares the mean values of variables between respondents who have daughters, those who have sons, and those who do not have children. The key variable, GRAND DAUGHTERS (or GRAND DAUGHTER DUMMY), would suffer endogenous biases if the grandchild's gender depends on grandparents' preference. If there is no difference between them, these groups can be comparable and balanced. Therefore, the estimation model presented in the following section is valid. Some respondents had both daughters and sons. Hence, the sample of respondents who have grandchildren is divided into three groups. The results of the F-test did not show significant differences in any variables between the three groups. Further, using a sample excluding the group of grandchildren of both genders, the results of the t-test also did not show significant differences in any variables. This indicates that the genders of grandchildren are exogenously determined and do not suffer from endogenous biases.

Table 3 demonstrates the mean difference test about VOTE FEMALE. Rate of voting for female candidates is 35% for respondents with granddaughters, and 28% for those without granddaughters. The difference is statistically significant at the 1% level. In other comparisons, the differences are not statistically significant. This is congruent with *Hypothesis 2*, but not *Hypothesis 1*. However, these results possibly suffer omitted variables biases and selection biases. For a closer examination of the Hypotheses, regression models should be used.

## 4. EMPIRICAL METHODOLOGY

The baseline model assesses the influence of family members on grandparents' voting behavior. The estimated function takes the following form:

VOTE FEMALE $_i$ = $\alpha_0$ + $\alpha_1$ GRAND DAUGHTERS $_i$ (or GRAND DAUGHTER DUMMY $_i$) +

$\alpha_2$ GRAND SONS $_i$ (or GRAND SON DUMMY $_i$) + $\alpha_3$ DAUGHTERS (or DAUGHTER DUMMY) + $\alpha_4$ SONS (or GRAND SON DUMMY) $_i$ + Xi B + u $_i$.

VOTE FEMALE $_i$ is the dependent variable, which is a dummy variable with 0 or 1; thus, the probit model was used. The key independent variable is GRAND DAUGHTERS to test the main hypothesis concerning the effect of having a granddaughter on voting behavior. Parents have an incentive to have an additional child if they prefer a son but cannot have a boy as the first child. There is a possibility that the number of granddaughters depends on parents' gender preference regarding children. Further, parents' preferences are possibly inherited from their parents (grandparents of the child) (Albanese et al., 2016). This results in an endogenous bias. Therefore, as an alternative specification, GRAND DAUGHTERS is replaced by the GRAND DAUGHTER DUMMY. Apart from granddaughters, I also examine the effects of having grandsons to consider the gender differences of grandchildren. Further, the number of daughters and sons are included to re-examine their effects on parents, following previous studies (Glynn & Sen, 2014; Milyo & Schosberg, 2000; Washington, 2008; Oswald & Powdthavee, 2010). In the same way as variables are applied to capture the presence of granddaughter, dummy variables are incorporated in the alternative specification. From *Hypotheses 1* and *2*, the predicted coefficients of DAUGHTERS (or DAUGHTER DUMMY ) and GRAND DAUGHTERS (or GRAND DAUGHTER DUMMY ) are positive. To test the opposite gender effects, GRAND SONS, GRAND SON DUMMY, SONS, and SON DUMMY are included. As is explained when the results of Table 2 are interpreted, some respondents have both genders of children and grandchildren. That is, GRAND DAUGHTER DUMMY and GRAND SON are not mutually exclusive. Hence, GRAND DAUGHTER DUMMY and GRAND SON DUMMY can be estimated even if they are included at the same time.

In addition, the vector of the control variables is denoted by $X_i$ and the vector of the estimated coefficients is denoted by B. Probability of voting for female candidates depends on the number of female candidates in candidate lists (Bonomi et al., 2013). As control variables, I added the number of female candidates (FEMALE CANIDAT) to the respondent's election district. To capture basic demographic and economic conditions, I incorporated UNIV, AGE, 17 income dummies, and 19 occupation dummies. Apart from the key variables introduced, family composition is an important factor that affects views on social and economic issues (Borrel-Porta et al., 2019; Oswald & Powdthavee, 2010; Washington, 2008). Therefore, the number of siblings (SISTERS and BROTHERS) and dummies for marital status are included separately. The estimation results for some of the control variables are not reported. However, all variables are included in all estimations.

Selection biases possibly occur, and therefore, should be controlled by the Heckman-type probit model, where the probit model is employed not only in the first stage but also in the second stage. In the first stage, independent variables should affect the possibility of having grandchildren. Hence, the youngest child is used because the child can become a parent after becoming an adult. Further, the number of daughters and sons are included because a larger number of children leads to a higher probability of respondents becoming grandparents. Further, respondents' ages, educational backgrounds, and gender are included. In the second stage, the set of control variables is the same as those in Table 4.

## 5. ESTIMATION RESULTS

### 5.1. Probit model

The results in Table 4 are calculated based on the sub-sample; the sample included the election district with female candidates, and respondents voted in the election. Furthermore, it is limited to those over 40, 50, and 60 years. In columns (1)–(3), the key variables are the number of daughters and granddaughters to test *Hypotheses 1* and *2*. In columns (4)–(6), instead, dummies are included to avoid endogenous biases. The coefficients of DAUGHTERS and DAUGHTER DUMMY show a positive sign, except in column (2). However, statistical significance was not observed in any of the results. Hence, *Hypothesis 1* was not supported. GRAND DAUGHTERS and GRAND DAUGHTER DUMMY yielded the expected positive sign, while being statistically significant in all results. This is consistent with *Hypothesis 2*. As for the presence of sons and grandsons, the coefficients were negative in most results, although statistical significance was not observed in any results. As for control variables, UNIV shows a significant negative sign in four of the six, indicating that highly educated individuals are less likely to vote for female candidates. In addition, most of the results did not show statistical significance.

For a robustness check of the results indicated in Table 4, I limited the sample to those who have children in Table 5 and those who have children over 30 years of age, at least in Table 6. The set of independent variables is equivalent to that in Table 4. However, the results of key variables to examine the effects of granddaughters and grandsons are presented. GRAND DAUGHTERS and GRAND DAUGHTER DUMMY present a significant positive sign in all columns, whereas GRAND SONS and GRAND SON DUMMY indicate a negative sign in all columns, despite being statistically insignificant. These results are in line with the results in Table 4. In Table 6, the signs of the variables are the same as those in Table 5. Furthermore, the statistical significance of GRAND DAUGHTERS increased to 1%, and GRAND DAUGHTER DUMMY increased to 5% level in columns (4) and (5) and 1% level in column (6). In addition, GRAND SONS and GRAND SON DUMMY become statistically significant when the sub-sample is limited to those over 60 years of age, indicating that those who have grandsons are less likely to vote for female candidates, and more likely to vote for male candidates. The marginal effects of GRAND DAUGHTERS are 0.05-0.07, implying that an additional daughter leads to 5-7% increase in the probability of voting for female candidates. Meanwhile, those of GRAND DAUGHTER DUMMY are 0.11–0.12, implying that respondents who have granddaughters are more likely to vote for female candidates by 11-12% compared to those who do not have granddaughters. The results on the effect of having granddaughters are robust to specifications and sub-samples.

In Tables 4-7, it is possible that the results suffer from selection biases caused by decision making about whether respondents have grandchildren. In comparison with decision-making about having children, selection biases are less likely to occur. However, for closer examination, I employ the Heckman-type two-step probit model, and the results are interpreted in the following sub-section.

**5.2. Heckman type two-step Probit model**

Tables 7 and 8 present the results for controlling selection biases using the Heckman-type two-step probit model. In the first step, I select respondents who have grandchildren. In Table 7, AGE_CHILD, DAUTHERS, SONS, and AGE show a positive sign, while being statistically significant at the 1% level. Children's age is related to the probability of having grandchildren.

This implies that people have children when they are at childbearing ages. The larger the number of children, their parents are more likely to have a grandchild. Older people are more likely to have grandchildren because their children are more likely to be of childbearing age. Meanwhile, a significant negative sign was observed for UNIV. In my interpretation, highly educated people place more importance on the quality of children than the quantity of children, leading to a smaller number of children. The opportunity cost of childrearing is higher for them compared to low-educated people. Inevitably, highly educated people are less likely to have grandchildren.

In the second step of Table 7, similar to Tables 4-6, I noted a positive sign for GRAND DAUGHTERS and GRANDDAUGHTER DUMMY, and their statistical significance. The absolute values of the marginal effects of GRAND DAUGHTERS are 0.04, implying that an additional granddaughter increases the probability by 4% that respondents voted for female candidates. The absolute values of GRAND DAUGHTER DUMMY are 0.10, implying that respondents with granddaughters are more likely to vote for female candidates by 10% compared to those without. The variables of grandsons did not indicate statistical significance at all. As shown in Table 8, the sample is further limited to respondents over 50 years old, because people below 50 years of age are unlikely to be grandparents. These results are similar to those in Table 7. However, the statistical significance of GRAND DAUGHTERS and GRANDDAUGHTER DUMMY reached the 1% level in columns (1), (3), and (4), while the absolute values of the marginal effects are almost the same as in Table 7.

The combined estimation results of Tables 4-8 about the effects of granddaughters are robust and strongly supported *Hypothesis 2*. The results of daughters did not support *Hypothesis 1*. Therefore, the effect of the granddaughter is more significant and sizable compared to that of the daughter. This can be interpreted in various ways. For instance, in an aging and male-dominated society, women are required to balance the demands of work and parenting. It is necessary to devise countermeasures against the falling birth rate and development of an environment in which parents can balance work and child-rearing. Beaman et al. (2009) found that exposure to a female political leader weakens stereotypes about gender roles in public and domestic spheres. To support working mothers and realize work-life and work-childrearing balance, female political leaders have a critical role to play in changing social norms and outdated structures. The findings of this study provide new evidence that grandparents have an incentive to support their granddaughters by improving their future environments.

Cross-gender effects were observed in father-daughter relations (Washington, 2008; Oswald & Powdthavee, 2010; Cronqvist & Yu, 2017). In the nineteenth century, dramatic improvements occurred in the legal rights of married women. As argued by Doepke and Tertilt (2009), men preferred women to have legal rights because men care about their own daughters, and because an expansion of women's rights increases educational investments in children. In other words, men relinquished some of their power to favor their own daughters once the importance of human capital increased. Finally, I test whether cross-gender effects were observed in the grandfather-granddaughter relationship. I add cross terms between the male dummy and key variables to the specifications in Tables 7 and 8. The results of the cross-terms are presented in Table 9. I did not observe statistical significance in any of the results. Therefore, there are no cross-gender effects on voting behavior between grandfathers and granddaughters, that is, the presence of granddaughters gives equal incentive to grandfathers and grandmothers to vote for female candidates.

6. CONCLUSION

In an unprecedented aging society, the grandparent-grandchild relationship becomes more crucial than ever before. Extending the existing works considering the effect of daughters on her parents' views and behaviors (Milyo & Schosberg, 2000; Oswald & Powdthavee, 2010; Glynn & Sen, 2014; Cronqvist & Yu, 2017), I analyzed how the presence of granddaughter influences her grandparents' voting behavior. The advantage of this study over previous literature is that it considers grandparents—since grandparents are less likely to influence their children's decision-making about having grandchildren than their own decision making about having children. Therefore, it is assumed that the presence of grandchildren and their genders are considered exogenous, in comparison with the presence of children.

I independently collected individual-level data directly after the election in Japan. Using the data, the major finding is that having a granddaughter leads the grandparents to vote for female candidates in the election by around 10% more, compared to not having one. However, this tendency was not observed for those with a grandson. The female social position is remarkably lower compared to the male social position in Japan, which provides a greater motivation to favor a granddaughter than grandson. Further, in contrast to existing works (Washington, 2008; Oswald & Powdthavee, 2010), having a daughter did not influence grandparents' voting behavior. The fact that the granddaughter's effect is far more significant than the daughter's effect reflects that granddaughter is too young to vote, while daughters can vote on their own will. Thus, grandparents voted for the future benefit of their granddaughters. Female political leaders are more likely to be beneficial to the interests of children (Carruthers & Wanamaker, 2015; Miller, 2008) compared to male political leaders. Grandparents decide between voting for their own interests or their grandchildren's. In my interpretation of the results, grandparents put more importance on granddaughters' lives in the future than to the remainder of their own lives. The contribution of this study is to indicate that grandparents voted as if they served as agents for their granddaughters.

Like all empirical studies, this study has some limitations. It did not analyze how the presence of granddaughters influences grandparents' views about specific policies. Further, inconsistent with existing works (Washington, 2008; Oswald & Powdthavee, 2010), cross-gender effects are not explored, that is, the effect of granddaughters did not differ between grandfathers and grandmothers. Therefore, it is necessary to further analyze the reasons for this. Additionally, access to ultrasound technologies and abortion services may alter this fact by allowing grandparents to choose the gender of their grandchildren, even though grandparents' preference is unlikely to be influential. However, these limitations do not undermine the value of this study. In contrast, they highlight that further work on the effects of the gender of grandchildren on voting behavior is warranted.

TABLE 1. Basic statistics and definition of variables used in estimation

|  | Definition | Mean | Max | Min |
|---|---|---|---|---|
| VOTE FEMALE | It is 1 if respondents vote for female candidate, 0 if they vote for male candidate. | 0.29 | 1 | 0 |
| GRAND DAUGHTERS | Number of respondent's granddaughters | 0.17 | 6 | 0 |
| GRAND SONS | Number of respondent's grandsons | 0.18 | 6 | 0 |
| GRAND DAUGHTER DUMMY | It is 1 if respondents have granddaughters, otherwise 0. | 0.11 | 1 | 0 |
| GRAND SON DUMMY | It is 1 if respondents have grandsons, otherwise 0. | 0.11 | 1 | 0 |
| DAUTHERS | Number of respondent's daughters | 0.65 | 4 | 0 |
| SONS | Number of respondent's sons | 0.69 | 4 | 0 |
| FEMALE CANDIDAT | Number of female candidates in a respondent's election district | 3.02 | 7 | 1 |
| UNIV | It is 1 if respondent graduated from university, otherwise 0. | 0.56 | 1 | 0 |
| INCOME | Respondent's annual household income (10 thousands yens) | 716 | 50 | 2000 |
| SISTERS | Number of respondent's sisters | 0.72 | 6 | 0 |
| BROTHERS | Number of respondent's brothers | 0.68 | 8 | 0 |
| AGE | Respondent's ages | 53.5 | 66 | 41 |
| FEMALE | It is 1 if respondent is female, otherwise 0 | 0.43 | 1 | 0 |
| MALE | It is 1 if respondent is male, otherwise 0 | 0.57 | 1 | 0 |
| AGE_CHILD | Ages of the youngest child among respondent's children | 21.6 | 63 | 0 |
| Observations |  | 3843 |  |  |

Note: The sample is restricted to respondents aged 41 years or older and residing in an election district where the number of female candidates is not 0. The sample used for AGE_CHILD is further restricted to respondents with children, leading to a reduction of 2618 candidates.

TABLE 2. Balance check

Difference test between groups of having only granddaughters, having only grandsons, and both of them.

| Variable | Grand-daughters (1) | Grandsons (2) | Both (3) | F-test Null-Hypothesis Ho:(1)=(2)=(3) | t-test Null-Hypothesis Ho:(1)=(2) |
|---|---|---|---|---|---|
| DAUTHERS | 0.97 | 1.03 | 1.15 | 2.34 (0.10) | 0.67 (0.50) |
| SONS | 1.11 | 1.12 | 1.21 | 0.96 (0.38) | 0.10 (0.92) |
| FEMALE CANDIDAT | 2.89 | 3.02 | 2.68 | 1.85 (0.16) | 0.66 (0.51) |
| UNIV | 0.48 | 0.42 | 0.38 | 1.81 (0.16) | 1.10 (0.27) |
| INCOME | 694 | 658 | 622 | 1.19 (0.30) | 0.68 (0.49) |
| SISTERS | 0.93 | 0.83 | 0.85 | 0.50 (0.60) | 0.92 (0.36) |
| BROTHERS | 0.74 | 0.88 | 0.89 | 1.15 (0.32) | 1.48 (0.14) |
| AGE | 60.8 | 61.2 | 66.7 | 1.95 (0.14) | 0.67 (0.51) |
| FEMALE | 0.52 | 0.57 | 0.58 | 0.76 (0.47) | 0.89 (0.37) |

Note: The sample is the same as that used in Table 1. The numbers in parentheses are p-values. Absolute t-values are reported in the t-test column.

TABLE 3. Mean difference test of VOTE FEMALE

| Grand-daughters (1) | Others (2) | t-test Null-Hypothesis Ho:(1)=(2) |
|---|---|---|
| 0.35 | 0.28 | 2.68*** (0.00) |
| Grand-sons (1) | Others (2) | t-test Ho:(1)=(2) |
| 0.32 | 0.29 | 1.46 (0.16) |
| Daughters (1) | Others (2) | t-test Ho:(1)=(2) |
| 0.30 | 0.28 | 1.27 (0.20) |
| Sons (1) | Others (2) | t-test Ho:(1)=(2) |
| 0.29 | 0.28 | 0.67 (0.50) |

Note: The sample is the same as that used in Table 1. Absolute t-values are reported in the t-test column.

*** indicates significance at the 1 % level.

TABLE 4. Results of the baseline specification Sub-sample divided by respondent's age (Probit model)

|  | (1) AGE>=40 | (2) AGE>=50 | (3) AGE>=60 | (4) AGE>=40 | (5) AGE>=50 | (6) AGE>=60 |
|---|---|---|---|---|---|---|
| GRAND DAUGHTERS | 0.03** (2.36) | 0.04*** (2.65) | 0.06** (3.48) | | | |
| GRAND SONS | −0.01 (−0.58) | −0.005 (−0.29) | −0.01 (−0.67) | | | |
| GRAND DAUGHTER DUMMY | | | | 0.06* (1.81) | 0.07** (2.24) | 0.09** (2.20) |
| GRAND SON DUMMY | | | | −0.02 (−0.73) | −0.01 (−0.48) | −0.01 (−0.35) |
| DAUTHERS | 0.01 (0.73) | −0.003 (−0.23) | 0.002 (0.10) | | | |
| SONS | 0.007 (0.57) | −0.007 (−0.49) | −0.02 (−0.56) | | | |
| DAUGHTER DUMMY | | | | 0.02 (1.16) | 0.002 (0.12) | 0.02 (0.56) |
| SON DUMMY | | | | 0.01 (0.50) | −0.02 (−0.83) | −0.01 (−0.26) |
| GRAND DAUGHTER DUMMY | 0.01 (0.58) | 0.01 (0.45) | 0.01 (0.55) | 0.01 (0.58) | 0.01 (0.45) | 0.01 (0.56) |
| UNIV | −0.03* (−1.86) | −0.04** (−1.98) | −0.02 (−0.37) | −0.04* (−1.90) | −0.04** (−2.01) | −0.02 (−0.47) |
| INCOME | −0.04* (−1.85) | −0.03 (−0.98) | −0.04 (−1.25) | −0.04* (−1.91) | −0.03 (−0.99) | −0.04 (−1.25) |
| SISTERS | −0.006 (−0.66) | −0.003 (−0.25) | −0.002 (−0.10) | −0.006 (−0.61) | −0.002 (−0.18) | 0.006 (0.04) |
| BROTHERS | −0.002 (−0.21) | −0.004 (−0.40) | 0.007 (0.44) | −0.001 (−0.16) | −0.003 (−0.36) | 0.007 (0.49) |
| AGE | 0.0004 (0.42) | 0.002 (0.81) | −0.02* (−1.75) | 0.0005 (0.53) | 0.002 (0.82) | −0.02* (−1.80) |
| FEMALE | 0.02 (0.99) | 0.03 (1.12) | 0.05 (1.02) | 0.02 (0.97) | 0.03 (1.08) | 0.04 (0.85) |
| Observations | 3,843 | 2,359 | 937 | 3,843 | 2,359 | 937 |
| Pseudo R-square | 0.01 | 0.01 | 0.03 | 0.01 | 0.01 | 0.03 |

Note: Numbers without parentheses are marginal effects. For convenience of interpretation, the marginal effects of INCOME are multiplied by 1000. Numbers in parentheses are z-values calculated using robust standard errors clustered in residential prefectures. **, and *** indicate significance at the 5%, and 1% levels, respectively. Various control variables are included, such as the respondent's marital status and job status dummies. However, these results have not been reported. The sample is respondents who resided in an election district where the number of female candidates is not 0.

TABLE 5. Results of baseline specification. Sub-sample which is limited to those who have their children and divided by respondent's ages. (Probit model)

|  | (1) AGE>=40 | (2) AGE>=50 | (3) AGE>=60 | (4) AGE>=40 | (5) AGE>=50 | (6) AGE>=60 |
|---|---|---|---|---|---|---|
| GRAND DAUGHTERS | 0.03** (2.09) | 0.03** (2.09) | 0.05** (2.45) | | | |
| GRAND SONS | −0.008 (−0.47) | −0.008 (−0.50) | −0.02 (−1.11) | | | |
| GRANDDAUGHTER DUMMY | | | | 0.06* (1.80) | 0.06* (1.87) | 0.07* (1.74) |
| GRAND SON DUMMY | | | | −0.02 (−0.55) | −0.02 (−0.62) | −0.02 (−0.57) |
| Observations | 2,627 | 1,784 | 768 | 2,627 | 1,784 | 768 |
| Pseudo R-square | 0.01 | 0.01 | 0.03 | 0.01 | 0.01 | 0.03 |

Note: Numbers without parentheses are marginal effects. Numbers in parentheses are z-values calculated using robust standard errors clustered in residential prefectures. **, and *** indicate significance at the 5%, and 1% levels, respectively. Neither DAUGHTERS nor SONS were included as independent variables. However, the results hardly change if DAUGHTERS or SONS are included. Apart from DAUGHTERS or SONS, the set of control variables included in the model are equivalent to those in Table 4. However, these results have not been reported. The sample is respondents who resided in an election district where the number of female candidates is not 0.

TABLE 6. Results of baseline specification. Sub-sample which is limited to those who have their children over 30 years old and divided by respondent's ages. (Probit model)

|  | (1) AGE>=40 | (2) AGE>=50 | (3) AGE>=60 | (4) AGE>=40 | (5) AGE>=50 | (6) AGE>=60 |
|---|---|---|---|---|---|---|
| GRAND DAUGHTERS | 0.05*** (2.75) | 0.05*** (2.71) | 0.07*** (2.61) | | | |
| GRAND SONS | −0.02 (−1.01) | −0.02 (−1.02) | −0.05* (−1.81) | | | |
| GRANDDAUGHTER DUMMY | | | | 0.11** (2.50) | 0.11** (2.46) | 0.12*** (2.67) |
| GRAND SON DUMMY | | | | −0.04 (−1.13) | −0.04 (−1.15) | −0.07* (−1.73) |
| Observations | 530 | 527 | 434 | 530 | 527 | 434 |
| Pseudo R-square | 0.04 | 0.04 | 0.05 | 0.04 | 0.04 | 0.05 |

Note: Numbers without parentheses are marginal effects. Numbers in parentheses are z-values calculated using robust standard errors clustered in residential prefectures. **, and *** indicate significance at the 5 %, and 1 % levels, respectively. Neither DAUGHTERS nor SONS were included as independent variables. However, the results hardly change if DAUGHTERS or SONS are included. Apart from DAUGHTERS or SONS, the set of control variables included in the model are equivalent to those in Table 4. However, these results have not been reported. The sample is respondents who resided in an election district where the number of female candidates is not 0.

TABLE 7. Results of baseline specification. (Heckman Probit model)

|  | (1) | (2) | (3) | (4) |
|---|---|---|---|---|
|  |  | Second-Step |  |  |
| GRAND DAUGHTERS | 0.04** (2.30) | 0.04* (1.72) |  |  |
| GRAND SONS | −0.003 (−0.14) | −0.001 (−0.03) |  |  |
| GRANDDAUGHTER DUMMY |  |  | 0.10** (2.56) | 0.10*** (2.67) |
| GRAND SON DUMMY |  |  | 0.01 (0.30) | 0.01 (0.29) |
| DAUTHERS |  | −0.02 (−0.51) |  | 0.002 (0.06) |
| SONS |  | −0.05 (−1.26) |  | −0.01 (−0.18) |
|  |  | First-step |  |  |
| AGE_CHILD | 0.07*** (10.5) | 0.07*** (10.3) | 0.07*** (10.3) | 0.07*** (10.5) |
| DAUTHERS | 0.50*** (7.81) | 0.50*** (7.82) | 0.50*** (7.82) | 0.50*** (7.77) |
| SONS | 0.41*** (6.44) | 0.41*** (6.43) | 0.41*** (6.43) | 0.41*** (6.46) |
| UNIV | −0.27*** (−3.83) | −0.27*** (−3.82) | −0.27*** (−3.82) | −0.27*** (−3.82) |
| AGE | 0.07*** (5.35) | 0.07*** (5.35) | 0.07*** (5.35) | 0.07*** (5.35) |
| FEMALE | −0.04 (−0.69) | −0.04 (−0.69) | −0.04 (−0.69) | −0.04 (−0.69) |
| Observations | 5025 | 5025 | 5025 | 5025 |
| Selected observations | 572 | 572 | 572 | 572 |
| Log pseudolikelihood | −1368 | −1368 | −1368 | −1368 |

Note: Numbers without parentheses are marginal effects. Numbers in parentheses are z-values calculated using robust standard errors clustered in residential prefectures. *, **, and *** indicate significance at the 10%, 5 %, and 1 % levels, respectively.

In the second stage, the set of control variables included in the model is equivalent to those in Table 4. However, apart from the key variables, the results are not reported. In the first stage, the results of all the independent variables are reported. The sample is respondents who resided in an election district where the number of female candidates is not 0.

TABLE 8. Results of baseline specification. Sub-sample which is limited to respondents over 50 years old. (Heckman Probit model)

|  | (1) AGE>=50 | (2) AGE>=50 | (3) AGE>=50 | (4) AGE>=50 |
|---|---|---|---|---|
|  | Second-Step | | | |
| GRAND DAUGHTERS | 0.04*** (2.64) | 0.04** (2.36) | | |
| GRAND SONS | 0.001 (0.04) | 0.002 (0.10) | | |
| GRANDDAUGHTER DUMMY | | | 0.10*** (2.85) | 0.09*** (3.07) |
| GRAND SON DUMMY | | | 0.02 (0.48) | 0.02 (0.47) |
| DAUTHERS | | −0.01 (−0.28) | | 0.01 (0.25) |
| SONS | | −0.04 (−1.07) | | −0.01 (−0.31) |
|  | First-Step | | | |
| AGE_CHILD | 0.07*** (9.25) | 0.07*** (9.25) | 0.07*** (9.25) | 0.07*** (9.25) |
| DAUTHERS | 0.56*** (7.92) | 0.56*** (7.96) | 0.56*** (7.94) | 0.56*** (7.92) |
| SONS | 0.48*** (6.18) | 0.48*** (6.19) | 0.48*** (6.17) | 0.48*** (6.18) |
| UNIV | −0.27*** (−3.87) | −0.27*** (−3.87) | −0.27*** (−3.87) | −0.27*** (−3.87) |
| AGE | 0.09*** (7.76) | 0.09*** (7.75) | 0.09*** (7.75) | 0.09*** (7.76) |
| FEMALE | −0.005 (−0.07) | −0.005 (−0.07) | −0.005 (−0.07) | −0.005 (−0.07) |
| Observations | 2459 | 2459 | 2459 | 2459 |
| Selected observations | 563 | 563 | 563 | 563 |
| Log pseudolikelihood | −1280 | −1280 | −1280 | −1280 |

Note: Numbers without parentheses are marginal effects. Numbers in parentheses are z-values calculated using robust standard errors clustered in residential prefectures. *, **, and *** indicate significance at the 10%, 5 %, and 1 % levels, respectively.

In the second stage, the set of control variables included in the model is equivalent to those in Table 4. However, apart from the key variables, the results are not reported. In the first stage, the results of all the independent variables are reported. The sample is respondents who resided in an election district where the number of female candidates is not 0.

TABLE 9. Results to test the cross-gender effect by adding cross terms. Sub-sample which is limited to respondents over 50 years old. (Heckman Probit model)

|  | (1) | (2) | (3) | (4) |
|---|---|---|---|---|
|  |  |  | AGE>=50 | AGE>=50 |
|  | Second-Stage | | | |
| GRAND DAUGHTERS*MALE | −0.01 |  | −0.01 |  |
|  | (−0.18) |  | (−0.15) |  |
| GRAND SONS*MALE | 0.03 |  | 0.03 |  |
|  | (0.76) |  | (0.84) |  |
| GRAND DAUGHTERS DUMMY *MALE |  | 0.08 |  | 0.08 |
|  |  | (0.93) |  | (1.13) |
| GRAND SONS DUMMY*MALE |  | 0.08 |  | 0.08 |
|  |  | (1.26) |  | (1.34) |
| Observations | 5025 | 5025 | 2459 | 2459 |
| Selected observations | 572 | 572 | 563 | 563 |
| Log pseudolikelihood | −1368 | −1368 | −1280 | −1280 |

Note: Numbers without parentheses are marginal effects. Numbers in parentheses are z-values calculated using robust standard errors clustered in residential prefectures. *, **, and *** indicate significance at the 10%, 5 %, and 1 % levels, respectively.

In the second stage, the set of control variables included in the model are equivalent to those in columns (1) and (3) of Tables 7 and 8, although the female dummy is replaced by the male dummy (MALE). However, apart from the key variables, the results are not reported. The sample is respondents who resided in an election district where the number of female candidates is not 0.

Endnotes:

[1] A firm's chief executive officer increased investment for corporate social responsibility after having a daughter (Cronqvist & Yu, 2017).

[2] The grandparents' caregiving has a significant effect on their health status (Ku et al., 2012; Reinkowski 2013; Di Gessa et al., 2016) and participation in social activities (Arpino & Bordone, 2017).

[3] Informal care of grandchildren reduced grandparents' subjective well-being (Brunello & Rocco, 2019). The negative effect is significantly larger for maternal grandmother compared to paternal one (Yamamura & Brunello, 2021).

[4] Female incumbents did not increase share of vote for new women candidates (Clayton & Tang, 2018). An additional female candidate's win leads to reduction of female candidates in the following elections (Kuliomina, 2018).

[5] Under the COVID-19 pandemic, even for two-income households, wives were far more likely to work from home to take care of their primary school children (Yamamura & Tsustui, 2021a), resulting in their worser mental health (Yamamura & Tsustui, 2021b).

[6] The former Tokyo Olympics chief, Yoshiro Mori, has drawn international criticism after his embarrassing gaffes on derogatory comments about women. Besides the condemnation from other countries, criticism increased to call for his resignation from Japanese citizens. Inevitably, he resigned.

[7] Data is extracted from the official website of the Statistics Bureau, Ministry of Internal Affairs and Communications http://www.soumu.go.jp/johotsusintokei/statistics/statistics05.html (accessed on April 5, 2018).

[8] In Japan, the voting age for national elections was 20 years and was lowered to 18 from the election of 2016.